\documentclass[12pt]{iopart}
\usepackage{iopams,epsfig,amsfonts,amssymb}  
\def\sqr#1#2{{\vcenter{\hrule height.#2pt\hbox{\vrule width.#2pt
height#1pt \kern#1pt \vrule width.#2pt}\hrule height.#2pt}}}
\def\hook{\hbox{\vrule height0pt width4pt depth0.3pt
\vrule height7pt width0.3pt depth0.3pt \vrule height0pt width2pt
depth0pt} }
\newcommand{\beqn}{\begin{equation}}
\newcommand{\eeqn}{\end{equation}}
\newcommand{\bbox}{\vrule height7pt width4pt depth1pt}

\newcommand{\sect}[1]{\setcounter{equation}{0}\bigskip\medskip
\section{#1}\smallskip}
\newcommand{\subsect}[1]{\medskip\subsection{#1}\smallskip}

\newtheorem{THEOREM}{Theorem}[section]  %chapter
\newenvironment{theorem}{\begin{THEOREM} \hspace{-.85em} {\bf :} 
}%
                        {\end{THEOREM}}
\newtheorem{PROPOSITION}[THEOREM]{Proposition}
\newenvironment{proposition}{\begin{PROPOSITION} \hspace{-.85em} 
{\bf :} }%
                            {\end{PROPOSITION}}
\newtheorem{DEFINITION}[THEOREM]{Definition}
\newenvironment{definition}{\begin{DEFINITION} \hspace{-.85em} {\bf 
:} \rm}%
                            {\end{DEFINITION}}
\newcommand{\thm}{\begin{theorem}}
\newcommand{\lem}{\begin{lemma}}
\newcommand{\pro}{\begin{proposition}}
\newcommand{\dfn}{\begin{definition}}
\newcommand{\ethm}{\end{theorem}}
\newcommand{\elem}{\end{lemma}}
\newcommand{\epro}{\end{proposition}}
\newcommand{\edfn}{\bbox\end{definition}}
\newcommand{\prf}{\noindent{\bf Proof:} }
\newcommand{\eprf}{\bbox}
\def\br{\begin{eqnarray}}
\def\er{\end{eqnarray}}
\def\brn{\begin{eqnarray*}}
\def\ern{\end{eqnarray*}}
\def\er{\end{eqnarray}}
\def\beq{\begin{equation}}
\def\eeq{\end{equation}}

\def\vt{\vartheta}

\def\d{\delta}

\def\r{\rho}
\def\A{\mathcal{A}}
\def\B{\mathcal{B}}

\def\L{\mathcal{L}}
\def\T{\mathcal{T}}
\def\C{\mathcal{C}}
\def\F{\mathcal{F}}
\def\E{\mathcal{E}}
\def\S{\mathcal{S}}
\def\X{\mathcal{X}}
\eqnobysec
\begin{document}
\title{Coframe energy-momentum current. Algebraic properties. }
\author{Yakov Itin}
\address{${}^{a}$ Institute of Mathematics,  
Hebrew University of Jerusalem\\
 Givat Ram, Jerusalem 91904, Israel, \\
${}^{b}$ Jerusalem College of Engineering,\\
Ramat Beit Hakerem, Jerusalem 91035, Israel\\
itin@math.huji.ac.il}

\begin{abstract}
The coframe (teleparallel) description of gravity is known as a viable alternative 
to GR. 
One of advantages of this model is the existence of 
a conserved energy-momentum current witch is covariant 
under all symmetries of the three-parameter Lagrangian. 
In this paper we study the relation between the covector valued 
current and the energy-momentum tensor. 
Algebraic properties of the conserved current for different values of 
parameters are derived. 
It is shown  that the tensor 
corresponding to the coframe current is traceless and, in contrast to the electromagnetic 
field, has in general a non vanishing antisymmetric part. 
The symmetric part is also non zero for all values of the parameters. 
Consequently, the conserved current involves the energy-momentum as well as 
the  rotational (spin) properties of the field.
\end{abstract}
Keywords: Teleparallel equaivalent of GR, conserved current, energy-momentum tensor.  
%--------------------------
\sect{Introduction. Coframe gravity}                  
%----------------------------

The {\it teleparallel} description of gravity has been studied for a long time. 
It has   recently evoked considerable  interest  for various reasons:
\begin{itemize}
\item[1.] This theory is a natural application of gauge principles to spacetime 
symmetries  \cite{H-N}-\cite{Kawai}.
\item[2.] The teleparallel Lagrangian constitutes a self consistent
sector of metric-affine gravity \cite{hehl95}, i.e., the gauge theory of the
4-dimensional affine group in the presence of a metric.
\item[3.] It represents a 1-parameter family of viable models of
gravity, all with the Schwarzschild solution for the spherically symmetric 1-body problem
\cite{H-S}, \cite{Hehl98}, \cite{itin1}.
\item[4.] In comparison with the standard GR the teleparallel theory has
an improved behavior of its Lagrangian insofar as it admits a covariantly
defined energy-momentum current \cite{Per}, \cite{itin2}.
\item[5.] The canonical analysis  of the teleparallel construction 
has remarkable advantages in comparison with that of standard GR 
\cite {Maluf3}, \cite{Blag2},  \cite{Blag1}. 
\item[6.] The  teleparallel technique was successfully applied  for a   
transparent treatment of  Ashtekar's complex variables \cite{Mielke} and 
for the  tensorial proof of the positivity of the energy in GR \cite{Nester}.
\end{itemize}
The teleparallel theory is usually considered in a tensorial representation. 
A frame and a non-symmetric teleparallel connection (via its torsion and
contortion tensors) play the role of the basic dynamical variables. 
The connection between these tensors is given by a constraint equation
which represents the vanishing of the Riemannian curvature. 
The teleparallel theory allows also for an alternative coframe
representation \cite{Hehl98}, \cite{itin2}, which follows 
the general construction of  metric-affine gravity  \cite{hehl95}.
In this representation the field equation and the conserved energy-momentum current  
turn out to be completely similar to those of the Maxwell-Yang-Mills theory. 
Such an analogy may be useful for the transformation of the Yang-Mills technique to 
gravity. 
Furthermore, it lays a common basic framework for gravitational 
and electromagnetic fields. 

Let us briefly recall  the coframe approach to gravity:\\
Consider a smooth coframe field $\{\vt^a(x), \ x\in M, \ a=0,1,2,3\}$
defined  on a differential manifold $M$. 
The coframe constitutes at every point $x\in M$ a set of four linearly
independent 1-forms, i.e., a basis of the cotangent vector space $T^*M_x$.
The 1-forms $\vt^a$ are considered to be pseudo-orthonormal.
This assumption fixes  a metric on $M$ which is represented by the
coframe as 
\begin{equation}\label{metr}
g=\eta_{ab}\vt^a\otimes\vt^b,
\end{equation}
where $\eta_{ab}=diag(-1,1,1,1)$ is the Lorentzian metric.
Thus, the coframe field $\vt^a$ plays the role of a dynamical variable.
This is in contrast to  metric gravity of GR based on the metric tensor $g$.
The dimension of the coframe 1-forms are of length in
accordance with the length square dimension of the metric (\ref{metr}). 
We will use the geometrized units  system where $G=c=\hbar=1$. \\
The most general Lagrangian 4-form for the coframe field (minimally coupled to matter)
that is quadratic in the first order derivatives is 
\begin{equation}\label{lagr}
\L=\frac 1{2} \C_a\wedge*\F^a+{}^{(mat)}\L,
\end{equation}
where $\C^a:=d\vt^a$ is a 2-form of the coframe field strength.
The 2-form $\F^a$   is a linear combination of three
2-forms
\begin{equation}\label{F-def}
\F^a:=\rho_1\C^a+\rho_2e^a\hook(\vt^m\wedge \C_m)+
\rho_3e_m\hook(\vt^a\wedge \C^m).
\end{equation}
For the representation of $\F_a$ via the irreducible pieces see \cite{hehl95}.
The free parameters $\rho_i$ are dimensionless. 
For instance, the set 
$$\rho_1=0, \qquad \rho_2+2\rho_3=0$$
represents  {\it the teleparallel equivalent of general relativity} - 
$GR_{||}$.
It is well known \cite{hehl95}, however,  that even the more general set
of parameters: 
$$\rho_1=0, \qquad  \rho_2-arbitrary, \qquad \rho_3 \ne 0$$ 
produces just a unique static spherically  symmetric 
coframe solution. It corresponds to the  Schwarzschild metric
\cite{itin1}.\\
The Lagrangian (\ref{lagr}) is manifestly  diffeomorphic invariant.
It is also invariant under global transformations of the
coframe $\vt^a\to {A^a}_b\vt^b$, where ${A^a}_b\in SO(1,3)$. 
The $GR_{||}$ is a unique  local 
Lorentz invariant teleparallel model, where the transformation matrix
${A^a}_b$ is permitted to be a function of a point $x\in M$.\\
The variation of the Lagrangian (\ref{lagr}) yields the field equation
of the Yang-Mills type
\begin{equation}\label{feq1}
d*\F_a=\T_a+{}^{(mat)}\T_a,
\end{equation}
where $\T_a$ is a covector-valued 3-form constructed  from the first
order derivatives of the coframe:
\begin{equation}\label{cur}
\T_a=(e_a\hook\C^m)\wedge *\F_m-e_a\hook \L.
\end{equation}
${}^{(mat)}\T_a$ represents the energy-momentum current of the material field. 
Again, this expression (\ref{cur}) is similar to the energy-momentum 
current of the Yang-Mills field. 
A straightforward consequence of the field equation (\ref{feq1}) is 
the conservation law 
\begin{equation}\label{cons}
d(\T_a+{}^{(mat)}\T_a)=0.
\end{equation}
Observe that, in contrast to  standard GR, the conserved value is the 
total current of the system, not the material current itself.  
Certainly, this situation is more physical. \\ 
The conserved current (\ref{cur}) is local, i.e., constructed from the
fields and their derivatives at a point. 
It is invariant under the
diffeomorphisms of the manifold and transforms as a covector under
global Lorentz transformations of the coframe. 
Certainly, it is not a tensor if local Lorentz transformations of
the coframe are applied. 
Note, however, that for general values of parameters the coframe
Lagrangian also does not invariant under such transformations.
Thus the current $\T_a$ obeys all the symmetries of the Lagrangian. 
It is proved \cite{itin2} to be related to the diffeomorphism invariance
symmetry of the Lagrangian (\ref{lagr}).
Consequently, (\ref{cur})  represents the energy-momentum current of the
coframe field. 

In the present paper  we study the algebraic properties
of the covector-valued 3-form current $\T_a$ and its relation to the
energy-momentum tensor.  The tensor corresponding to the current
(\ref{cur}) is shown to be traceless for all values of the parameters
$\rho_i$. This is similar to the energy-momentum current of the
electromagnetic field.  Correspondingly, the gravitons in all coframe
models are massless as in  standard GR. For the trace of the gauge
energy-momentum, see \cite{hehl95}.\\ 
It is proved, in contrast to
electromagnetic field,  that the tensor corresponding to the coframe
current has in general non vanishing symmetric and antisymmetric
parts.  By the Poincar{\'e} gauge theory of gravity the antisymmetric part
of the current is connected to the rotational (spin) properties of the
field \cite{hehl95}.  It is equivalent to a scalar valued 2-form
$\S=e_a\hook \T^a$. \\ 
The coframe current preserves the symmetries of
the Lagrangian for all values of the parameters except for the special
case $\rho_2+2\rho_3=0$, which corresponds to  standard GR. \\ 
We briefly discuss how the Einstein theory (in the coframe representation) 
is embedded in the family of viable coframe models. 
%------------------------------------------
\sect{Algebraic relations}
%------------------------------------------
In this section we describe some algebraic relations which will be useful
for a treatment of the coframe current (\ref{cur}). 
There is no real advantage in restricting to dimension 4 and to the Lorentzian signature, 
so we work in this section (only) on a manifold  of an arbitrary dimension and signature. 
%-------------------------
\subsect{Basic facts}
%-------------------------
Let an $n$-dimensional  manifold $M$ endowed with a coframe field
$\{\vt^a(x), \ x\in M, \ a=1,...,n\}$ be given.
The coframe is accepted to be "pseudo-orthonormal", i.e., the metric on $M$
is represented via the coframe as
\begin{equation}\label{gen-met}
g=\eta_{ab}\vt^a\otimes\vt^b,
\end{equation}
where $\eta_{ab}=diag(-1,...,-1,1,...,1)$ is the pseudo-Euclidean metric.
Let the number of negative entries in $\eta_{ab}$ be equal to $i$, 
i.e., the  signature of the manifold $M$ is $(i,n-i)$ (in the Lorentzian case $i=1$). 
We use in this paper the coframe index notation. 
All indexed objects are diffeomorphic invariants and global Lorentz covariants. 
The metric $\eta_{ab}$ and its 
inverse $\eta^{ab}$ will be used to lower and raise the indices.  

Denote by $\F(M)$ the algebra (commutative and associative) of 
functions on $M$ and
by $\X(M)$ the $\F(M)$-module of vector fields on   $M$. 
Let $\Omega^1(M)$ designates module of 1-forms dual to $\X(M)$, i.e., the set of  
$\F(M)$-linear maps $\X(M)\to\F(M)$. 
Denote by $\Omega^p(M)$ the module of differential 
$p$-forms on $M$.  

The inner product operation $\hook:\X(M)\times\Omega^p(M)\to\Omega^{p-1}(M)$
is linear in both operands and 
acts on an exterior product of forms by 
the modified Leibniz rule. Namely, for $\A\in \Omega^p(M), \B\in \Omega^q(M), 
X\in\X(M)$
\begin{equation}\label{inner}
X\hook(\A\wedge\B)=(X\hook\A)\wedge\B+(-1)^p\A\wedge(X\hook\B). 
\end{equation}
%It is enough to define the Hodge dual  
%map $*:\Omega^p\to \Omega^{n-p}$ by its action on the basis forms
%\begin{equation}\label{hodge}
%*(\vt^{a_1}\wedge\cdots\wedge\vt^{\a_p})=(-1)^k\vt^{a_{p+1}}\wedge\cdots\wedge\vt^{a_n}, 
%\end{equation}
%where $k$ is equal to the number of time indices in the set $\{a_{p+1},\cdots,a_n\}$.  
%The set $\{a_1,\cdots,a_p,a_{p+1},\cdots, a_n\}$ is taken to be 
%an even permutation of $\{1,\cdots, n\}$. 
%This definition is extended by linearity to an arbitrary differential form. 
The Hodge dual  map $*:\Omega^p(M)\to \Omega^{n-p}(M)$ is defined via the metric 
(\ref{gen-met}) in a standard manner \cite{hehl95}. 
Its square  is an identity operator (up to a sign): 
\begin{equation}\label{h-sq}
*^2\A=(-1)^{p(n-p)+i}\A.
\end{equation}
Two forms $\A$ and $\B$ of the same degree satisfy the commutative rule 
\begin{equation}\label{comm}
\A\wedge*\B=\B\wedge*\A.
\end{equation}
Denote by $e_a$ the basis vectors dual to $\vt^a$. 
The duality is expressed by the inner product operation as 
$e_a\hook\vt^b=\d^b_a$. 
Recall the useful formulas \cite{hehl95} which will be frequently 
used  subsequently. 
The relations 
\begin{equation}\label{h-w1}
e_a\hook*\A=*(\A\wedge\vt_a)
\end{equation}
and
\begin{equation}\label{h-w2}
\vt_a\wedge *\A=(-1)^{p+1}*(e_a\hook \A)
\end{equation}
hold for an arbitrary form $\A$.
Trivial consequences of the antisymmetry are 
\begin{equation}%\label{form}
\vt^a\wedge(\vt_a\wedge \A)=0,
\end{equation}
and 
\begin{equation}%\label{form2}
e_a\hook(e^a\hook \A)=0.
\end{equation}
The successive actions of $\hook$ and $\wedge$ operations obey the properties
\begin{equation}\label{form1}
\vt^a\wedge(e_a\hook \A)=p\A,
\end{equation}
and 
\begin{equation}\label{form2}
e_a\hook(\vt^a\wedge \A)=(n-p)\A.
\end{equation}

%------------------------
\subsect{Form - tensor equivalence}
%------------------------

Let a covector-valued  $(n \ \!{\textrm -}1)$-form $\T_a$   on $M$ be given. 
This object has $n^2$ independent components, exactly as a generic 
second-rank tensor. Certainly, the coincidence is not by chance. 
Consider the Hodge dual form $*\T_a$. 
This  covector-valued  $1$-form may be regarded as a map:
 \begin{equation}\label{map1}
*\T_a:\X(M)\to\Omega^1(M),
\end{equation}
i.e., for a given vector field $X\in\X(M)$, $*\T_a(X)$ are components of a 1-form. 
This 1-form may  be successively regarded as a map to functions on $M$:
 \begin{equation}\label{map2}
*\T_a(X):\X(M)\to\F(M).
\end{equation}
The composition of the maps (\ref{map1}) and (\ref{map2}) defines a map  
 \begin{equation}\label{map3}
T:\X(M)\times\X(M)\to\F(M),
\end{equation}
which represents a $(2,0)$-rank tensor $T(X,Y)$.  
Relative to the basis $e_a$ this tensor obtains the component-wise 
form  $T_{ab}:=T(e_a,e_b)$, which is 
similar to the ordinary coordinate-wise notation of a  second-rank tensor. 
It should be noted, however, that the components $T_{ab}$ are invariants under 
a coordinate transformation. 
They constitute a tensor under the global pseudo-orthonormal transformations 
of the coframe $\vt^a$ (and the corresponding transformations of the basis 
vectors $e_a$).  \\
The equivalence between the current $\T_a$ and the tensor $T_{ab}$ 
is given by 
\begin{equation}\label{cur-ten}
\T_a=T_{ab}*\vt^b,
\end{equation}
i.e., tensor is represented as a matrix of the coefficients
of the current in the odd basis $*\vt^a$ of $\Omega^3$. 
Observe that two sides of (\ref{cur-ten}) are odd 3-forms, thus, $T_{ab}$ is an even 
0-form. 
Invert (\ref{cur-ten}) by applying (\ref{h-sq}) and (\ref{h-w1}) 
to obtain the explicit expression 
\begin{equation}\label{ten-cur}
T_{ab}=(-1)^{n-1+i} \ e_b\hook *\T_a=(-1)^{n-1+i}*(\T_a\wedge \vt_b).
\end{equation} 
Instead of the coframe representation (\ref{cur-ten}) an alternative coordinate 
representation of the current $\T_a$ can be introduced by $\T_a=T_{a\mu}*dx^\mu$. 
The object $T_{a\mu}$ is a covector relative to coordinate transformations and a 
covector  relative to coframe transformations. It is not, however, a tensor. 
Only on a flat manifold, where a closed pseudo-orthonormal coframe can be 
defined by $\vt^a=dx^a$, the distinction between the indices is disappeared and 
  $T_{a\mu}$ turns to be a tensor. \\
An alternative current-tensor relation (for electromagnetic field) was recently proposed 
by Hehl and Obukhov \cite{HO}, \cite{book}. 
In their approach the equivalence is defined by the relation
\begin{equation}\label{H-O1}
{T_a}^b=\diamond(\vt^b\wedge\T_a),
\end{equation}
where $\diamond$ denotes the dual with respect to the Levi-Civita density. 
The relation (\ref{H-O1}) is, certainly, in a more general setting than (\ref{cur-ten}),
(\ref{ten-cur}), because it allows to manage a current defined on a manifold without 
metric, i.e., for coframes transformed under $GL(n)$. 
Such framework is basic  for Hehl-Obukhov axiomatic construction of 
electromagnetic theory. 
In our pseudo-orthonormal coframe approach to gravity the metric and the Hodge 
dual are defined from the beginning. \\
On the other hand the relation (\ref{H-O1}) defines the $(1,1)$-type tensor and in order to consider its symmetric properties the contraction with some  metric tensor  
have to be taken also here. 
It should noted, however, that the trace of the tensor can be extracted irreducibly 
under $GL(n)$ already in (\ref{H-O1}). 

%------------------------
\subsect{Irreducible decomposition  of a current}
%------------------------
A covector-valued $(n \ \!{\textrm -}1)$-form admits an irreducible 
decomposition under the action of the pseudo-orthonormal group $SO(i,n-i)$. 
For a given $(n \ \!{\textrm -}1)$-form $\T_a$ define two pseudo-orthonormal 
(and diffeomorphic) invariants: an $n$-form 
\begin{equation}\label{tr-def}
\T=\vt^a\wedge \T_a,
\end{equation}
 and an $(n \ \!{\textrm -}2)$-form 
\begin{equation}\label{tsp-def}
\S=e_a\hook \T^a. 
 \end{equation}
The  scalar-valued $n$-form $\T$  
is an invariant linear combination of the components 
of the tensor $T_{ab}$. 
The only such invariant of a tensor is its trace. 
\pro 
The $n$-form $\T=\vt^a\wedge \T_a$ of an arbitrary  vector-valued 
$(n \ \!{\textrm -}1)$-form $\T_a$ satisfies the relation 
\begin{equation}\label{trace1}
\T={T^a}_a*1,
\end{equation}
where ${T^a}_a=\eta^{ab}T_{ab}$ is the trace of the tensor.
\epro
\prf
Insert the definition (\ref{cur-ten}) into (\ref{trace1}) to obtain
\begin{equation}\label{pr1-1}
\vt^a\wedge\T_a=T_{ab}\vt^a\wedge *\vt^b
\end{equation}
Use the  relation (\ref{h-w2}) to get 
$\vt^a\wedge *\vt^b=\eta^{ab}*1$. 
Thus  (\ref{pr1-1}) yields (\ref{trace1}). \\
\eprf\\
The  scalar valued $(n \ \!{\textrm -}2)$-form $\S$ has 
$n(n-1)/2$ independent components exactly as a generic  
 antisymmetric tensor.
\pro
The relation 
\begin{equation}\label{asym}
\S=-T_{[ab]}*(\vt^a\wedge \vt^b)
\end{equation}
holds for an arbitrary vector-valued $(n \ \!{\textrm -}1)$-form $T^a$.
\epro
\prf
Insert (\ref{cur-ten}) into the LHS of the (\ref{asym}) and use 
(\ref{h-w1}) to obtain
\begin{equation}\label{pr3-1}
e_a\hook \T^a=T_{ab}e^a\hook *\vt^b=T_{ab}*(\vt^b\wedge \vt^a). 
\end{equation}
\eprf
\pro
The irreducible decomposition of a covector-valued  $(n \ \!{\textrm -}1)$-form $\T^a$ 
under the (pseudo) orthonormal group is 
\begin{equation}\label{decomp}
\T_a={}^{(sym)}\T_a+{}^{(ant)}\T_a+{}^{(tr)}\T_a,
\end{equation}
where the trace part is 
\begin{equation}
{}^{(tr)}\T_a=\frac 1n \ e_a\hook \T,
\end{equation}
the antisymmetric part is
\begin{equation}\label{ASTM}
{}^{(ant)}\T_a=\frac 12 \ \vt_a\wedge \S,
\end{equation}
and the symmetric traceless part is 
\begin{equation}\label{ant}
{}^{(sym)}\T_a=\T_a-\frac 1n \ e_a\hook\T -\frac 12 \ \vt_a\wedge \S.
\end{equation}
\epro
\prf 
Let us express the irreducible parts of the current via the tensor components $T_{ab}$.
Using (\ref{trace1}) the trace part takes the form 
\begin{equation}\label{pr1-4}
{}^{(tr)}\T_a=\frac 1n {T^m}_m*\vt_a.
\end{equation}
As for the antisymmetric part by (\ref{asym})
\begin{equation}\label{pr1-5}
{}^{(ant)}\T_a=T_{[ab]}*\vt^b.
\end{equation}
Finally, the traceless symmetric part is 
\br\label{pr1-6}
{}^{(sym)}\T_a&=&T_{ab}*\vt^b-T_{[ab]}*\vt^b-\frac 1n {T^m}_m*\vt_a\nonumber\\
&=&(T_{(ab)}*\vt^{b}-\frac 1n \eta_{ab}{T^m}_m)*\vt^{b}.
\er
The irreducible decomposition of the current (\ref{decomp}) can now be regarded as
 \begin{equation}\label{decomp1}
T_{ab}=(T_{(ab)}-\frac 1n \eta_{ab}{T^m}_m)+T_{[ab]}+\frac 1n \eta_{ab}{T^m}_m,
\end{equation}
i.e., as an ordinary decomposition of a second rank tensor to a trace, 
antisymmetric, and symmetric traceless parts. 
\eprf
%------------------------
\subsect{Quadratic relations} 
%------------------------
The structure of the coframe current (\ref{cur})  is similar to the 
Yang-Mills-Maxwell energy-momentum current. 
Both expressions are quadratic in first order derivatives of the corresponding fields. 
Consider a scalar-valued $p$-form $\A$  which represents a generalized 
strength of a model based on a $(p-1)$-form field. 
In an analogy to (\ref{cur}), the $(n-1)$-form of current 
for this field is of the form 
\begin{equation}\label{z-sq0}
\tilde{\T_a}=(e_a\hook \A)\wedge *\A-\frac 12 e_a\hook(\A\wedge*\A).
\end{equation} 
The symmetry of the tensor corresponding to this current is governed 
by a scalar valued $(n-2)$-form
\begin{equation}\label{z-sq1}
\tilde{\S}=(e_a\hook \A)\wedge (e_a\hook*\A).
\end{equation}
As it will be shown in the consequence,  the vanishing of this form guarantees  
the symmetry of the corresponding tensor. 
Let us consider the $(n-2)$-form $\tilde{\S}$ in a general setting (without connection 
to some quadratic Lagrangian). 
\thm
The $(n-2)$-form $\tilde{\S}$ is vanishing for an arbitrary $p$-form $\A$, i.e., 
\begin{equation}\label{z-sq}
(e_a\hook \A)\wedge(e^a\hook *\A)=0.
\end{equation}  
\ethm
\prf
Suppose the LHS of (\ref{z-sq}) is nonzero. 
So, the $n$-form
\begin{equation}\label{pr7-1}
B^{mn}:=\vt^m\wedge(e_a\hook \A)\wedge\vt^n\wedge(e^a\hook *\A)
\end{equation}   
is antisymmetric for the permutation of the indices $m$ and $n$.
Use (\ref{inner}) to rewrite  (\ref{pr7-1}) as
\begin{equation}\label{pr7-2}
B^{mn}=\Big(\d^m_a \A-e_a\hook(\vt^m\wedge \A)\Big) \wedge
\Big(\eta^{an}*\A-e^a\hook (\vt^n\wedge *\A)\Big)
\end{equation}
Thus, $B^{mn}$ is expressed as a sum  of four terms. The first one is: 
$$
\d^m_a \A \wedge\eta^{an}*\A=\eta^{mn}\A\wedge*\A\qquad {\textrm - symmetric.}
$$
The second term is 
\brn
&&-e^n\hook(\vt^m\wedge \A) \wedge *\A=-\A\wedge *\Big(e^n\hook(\vt^m\wedge \A)\Big)\\
&&\qquad=(-1)^{k_1} (\vt^n \wedge \A)\wedge *(\vt^m\wedge \A)\qquad {\textrm - symmetric},
\ern
where the value of the integer $k_1$ is defined by (\ref{h-w2}) and (\ref{h-sq}).\\
The third term is 
\brn
&&- \A\wedge e^m\hook (\vt^n\wedge *\A)=
-*\Big(e^m\hook (\vt^n\wedge *\A)\Big)\wedge *\A\\
&&\qquad =(-1)^{k_2}(\vt^m\wedge *\A)\wedge *(\vt^m\wedge *\A)
\qquad {\textrm - symmetric}.
\ern
Finally, the fourth term is 
\brn
&&\Big(e_a\hook(\vt^m\wedge \A)\Big) \wedge \Big(e^a\hook (\vt^n\wedge *\A)\Big)\\
&&=
(-1)^{k_1}*\Big(\vt_a\wedge*(\vt^m\wedge \A)\Big) 
\wedge *\Big(\vt^a\wedge*(\vt^n\wedge *\A)\Big)\\
&&=(-1)^{k_2}\vt^a\wedge*(\vt^n\wedge *\A)\wedge
\vt_a\wedge*(\vt^m\wedge \A)=0
\ern
Therefore $B^{mn}$ is symmetric. The contradiction proves that the LHS 
of (\ref{z-sq}) is zero. 
\eprf\\
Observe that (\ref{z-sq}) is trivial for $\A$ be a wedge product of basis forms. 
The nonlinearity of the relation, however, seems to put an obstacle 
to restrict the proof.\\
In the case of a strength $\A$ to be a vector-valued $p$-form the 
corresponding current involves the term 
$(e_a\hook \A)\wedge*\B$, where $\B$ is a $p$-form constructed from 
$\A$ by inner and wedge product with basis forms. 
Although in this case the term $(e_a\hook \A)\wedge(e^a\hook *\B)$ is not 
zero, in general, some useful relations involving this expression may be established. 
\pro
Two forms $\A$ and $\B$ of the same degree satisfy 
\begin{equation}\label{al-bet}
(e_a\hook \A)\wedge(e^a\hook *\B)=-(e_a\hook \B)\wedge(e^a\hook *\A). 
\end{equation} 
\epro
\prf
It is enough to open the brackets in the relation
$$
\Big(e_a\hook (\A+\B)\Big)\wedge\Big(e^a\hook *(\A+\B)\Big)=0.
$$
\eprf
\pro
A form $\A$ of the degree $n$ in $2n$-dimensional space satisfies
\begin{equation}\label{*-*}
(e_a\hook *\A)\wedge(e^a\hook *\A)=(-1)^{n+i+1}(e_a\hook \A)\wedge(e^a\hook \A)
\end{equation} 
\epro
\prf
It is enough to replace $\B$ by $*\A$ in (\ref{al-bet}).
\eprf\\
Certainly the two sides of (\ref{*-*}) are nonzero only for an odd $n$, i.e., 
in the dimensions: $2,6,10,$ etc. \\
For the vector field theory the strength is a form of the second degree. 
In this case the following useful rule to deal with the Hodge star is valid.  
\thm
Two forms $\A$ and $\B$ of the second degree satisfy 
\begin{equation}\label{al-bet*}
(e_a\hook \A)\wedge(e^a\hook *\B)=*\Big((e_a\hook \A)\wedge(e^a\hook \B)\Big)
\end{equation} 
\ethm
\prf
Write down the 2-form $\A$ via the components relative to the 
pseudo-orthonormal basis $\A=1/2A_{mn}\vt^m\wedge \vt^n$. 
Consequently $e_a\hook\A=A_{am}\vt^m$. 
Compute the LHS of (\ref{al-bet*}) 
\brn
(e_a\hook \A)\wedge(e^a\hook *\B)&=& A_{am}\vt^m\wedge(e^a\hook *\B)\\
&=&A_{am}\vt^m\wedge*(\B\wedge \vt^a)
\qquad \qquad\qquad \qquad {\textrm {by} \  (\ref{h-w1})} \\
&=&(-1)^iA_{am}*^2\Big(\vt^m\wedge*(\B\wedge \vt^a)\Big)
\qquad \quad {\textrm {by} \  (\ref{h-sq})}\\
&=&A_{am}*\Big(e^m\hook(\B\wedge \vt^a)\Big)
\qquad \quad \qquad \qquad{\textrm {by} \  (\ref{h-w2})}\\
&=&A_{am}*\Big((e^m\hook\B)\wedge \vt^a+\B\eta^{am}\Big)\\
&=&A_{am}*\Big((e^m\hook\B)\wedge \vt^a\Big)\\
&=&*\Big((e^m\hook\B)\wedge A_{am}\vt^a\Big)\\
&=&*\Big((e_a\hook \A)\wedge(e^a\hook \B)\Big)
\ern
\eprf\\
Observe that two sides of the equation (\ref{al-bet*}) are 
$(n \ \!{\textrm -}2)$-forms. 
In the case $\A=\B$ the RHS is zero as a wedge square of a 1-form. 
Certainly, it is a special case of (\ref{z-sq}).
%------------------------
\sect{Coframe current}
%------------------------
The form-tensor equivalence described above allows to study the  algebraic 
properties of the current in parallel to the corresponding algebraic properties 
of the tensor. 
In this section we deal with the coframe current (\ref{cur}) defined on a 
$4D$-manifold of Lorentzian signature. 
%------------------------
\subsect{Traceless property}
%------------------------
The current  (\ref{cur}) involves free parameters $\r_i$. It is natural to look 
for which values of these parameters the corresponding tensor is traceless. 
By  (\ref{pr1-4}) the traceless tensor corresponds to a current satisfied 
$\T:=\vt_a\wedge\T^a=0$.
\pro
The coframe current (\ref{cur}) is traceless for an arbitrary  choice 
of the parameters $\r_1,\r_2,\r_3$.
\epro
\prf
Calculate
\begin{equation}\label{3-1}
\T^a\wedge\vt_a=-\vt^a\wedge (e_a\hook\C^m)\wedge *\F_m+
\vt^a\wedge e_a\hook \L
\end{equation}
Use the relation (\ref{form1}) to obtain
\begin{equation}\label{3-2}
\T^a\wedge\vt_a=-2\C_m\wedge *\F^m+4\L=0.
\end{equation}
\eprf\\
The coframe field equation (\ref{feq1}) incorporates the current 
(\ref{cur}) as a source term. 
This current is traceless for an arbitrary coframe field $\vt^a(x)$,  
even for this that  does not satisfy the field equation. 
Let us look  now how this traceless property  influences upon the algebraic features 
of the pure coframe field equation. 
Take the material Lagrangian to be zero. 
Thus from (\ref{lagr}) $\L=\frac 1{2} \C_a\wedge*\F^a$. 
Construct the exterior product in two sides of  (\ref{feq1}) to obtain
\begin{equation}\label{trace-eq1}
\vt_a\wedge d*\F^a=0,
\end{equation} 
or, equivalently,
\begin{equation}\label{trace-eq2}
d(\vt_a\wedge *\F^a)=2\L.
\end{equation} 
Thus the on-shell value of the Lagrangian is an exact form. 
Insert in the LHS of (\ref{trace-eq2}) the definition  (\ref{F-def}) 
of the strength $\F^a$ to obtain 
\begin{equation}\label{trace-eq3}
(\r_1-2\r_3) \ d(\vt_a\wedge *\C^a)=2\L.
\end{equation} 
Observe some conclusions from the equation (\ref{trace-eq3}) which valid for a pure 
coframe field in vacuum. 
\begin{itemize}
\item[i)] Consider the coframe model with parameters 
$\r_1=2\r_3, \r_2 \ {\textrm - arbitrary}$. 
For all solutions of 
the corresponding field equation the on-shell value of the Lagrangian  is zero.  
\item[ii)] Conversely, consider the models with zero on-shell value of the Lagrangian. 
Let $\r_1$ be different from $2\r_3$. 
By  (\ref{trace-eq3}) the 3-form  $\vt_a\wedge *\C^a$ is exact. 
Via the Poincar{\'e} lemma it means the (local) 
existence of a 2-form $A$ satisfying $dA=\vt_a\wedge *\C^a$. Consequently, 
for such models the 2-form $A$ is an integral invariant. 
\item[iii)] Let two different models (with different $\r$'s) have a 
joint solution $\vt^a(x)$. 
Thus the corresponding Lagrangians have the same on-shell value in both models,  
up to the coefficient $(\r_1-2\r_3)$.
\item[iv)] Consider the set of viable models: $\r_1=0, \r_3 \ne 0, \r_2 \ \textrm{- 
arbitrary}$. 
Let two models with different $\r_2$ have a joint solution $\vt^a(x)$. 
This solution have to satisfy 
 \begin{equation}\label{trace-eq4}
(\vt_a\wedge *\C^a)\wedge *(\vt_b\wedge *\C^b)=0,
\end{equation} 
i.e., the pseudo-norm of the 3-form $\vt_a\wedge *\C^a$ is zero. 
The well known solution of a such type is the Schwarzschild coframe 
which appears in all viable models (for all values of $\r_2$). 
This solution satisfies $\vt_a\wedge *\C^a=0$, thus also (\ref{trace-eq4}). 
\end{itemize}
%------------------------
\subsect{Symmetric property of the current}
%------------------------
The  coframe current (\ref{cur}) is formally similar to the 
electromagnetic energy-momentum current expression. 
The important distinctions are: \\
i) The coframe Lagrangian and the corresponding current involve  
the covector-valued strengths, 
while the electromagnetic theory is based on the scalar-valued strength.\\
ii) Two coframe strengths $\C^a$ and $\F^a$ are different 
for a generic choice of parameters.\\ 
Consequently, the tensor corresponding to the coframe  current has not 
to be symmetric in general. 
By (\ref{ASTM})  the tensor is symmetric if and only if the 
2-form $\S$ vanishes.  \\
Let us examine for which values of the 
parameters the coframe current produces a pure symmetric tensor. 
\pro
The 2-form $\S$ vanishes identically if and only if $\r_2=\r_3=0$. 
Consequently, for all viable models $\S$ and, correspondingly, the antisymmetric part 
of the tensor are non-zero.
\epro
\prf
Calculate the 2-form $\S$ for the coframe current (\ref{cur}) using the global 
$SO(1,3)$ covariants \cite{i-k} 
 \begin{equation}\label{spin2}
\S=-(e_a\hook \C_m)\wedge(e^a\hook *\F^m)=
*\Big((e_a\hook \F_m)\wedge(e^a\hook \C^m)\Big).
\end{equation} 
The formula (\ref{al-bet*}) was applied in the last equation. 
Use the definition (\ref{F-def}) of $\F_a$ to obtain 
\br\label{spin3}
\S&=& \r_1*\Big((e_a\hook \C_m)\wedge(e^a\hook \C^m)\Big)+\nonumber\\
&& \r_2*\bigg((e_a\hook \C^m)
\wedge(e^a\hook (\vt^k\wedge(e_m\hook \C^k)\big)\Big)\bigg)+\nonumber\\
&& \r_3*\bigg((e_a\hook \C_m)
\wedge\Big(e^a\hook \big(\vt_m\wedge (e_k\hook\C^k)\big)\Big)\bigg).
\er
The $\r_1$-term vanishes as a square of a 1-form (more generally, such type 
expressions are zero by (\ref{z-sq})). 
Calculate the $\r_2$-term using the component-wise expression
$\C^a=\frac 12 {C^a}_{bc}\vt^b\wedge\vt^c$. 
We obtain 
\br
\r_2*(\cdot\cdot\cdot)&=&
\r_2{C^m}_{a}C_{ams}*\Big(\vt^q\wedge \big(e_a\hook (\vt^k\wedge \vt^s)\big)\Big)\nonumber\\
&=&\r_2{C^{ma}}_{p}(C_{am}-C_{qma})*(\vt^p\wedge\vt^q)
\er
Use the 2-indexed $SO(1,3)$ covariants  
(see the Appendix) to rewrite this term as 
\begin{equation}\label{term1}
\r_2*(\cdot\cdot\cdot)=\r_2\Big({}^{(4)}\!A_{pq}+{}^{(5)}\!A_{pq}\Big)*(\vt^p\wedge\vt^q)
\end{equation}
As for the $\r_3$-term
\br
\r_3*(\cdot\cdot\cdot)&=&
\r_3{C^{ma}}_q{C^k}_{ks}*\Big(\vt^q\wedge \big(e_a\hook(\vt_m\wedge \vt^s)\big)\Big)\nonumber\\
&=&\r_3({C^m}_{mp}{C^k}_{kq}-{{C_{pq}}^a}{C^k}_{ka})*(\vt^p\wedge\vt^q)
\er
The first coefficient in this expression is symmetric thus 
\begin{equation}\label{term2}
\r_3*(\cdot\cdot\cdot)=-\r_3{{C_{pq}}^a}{C^k}_{ka}*(\vt^p\wedge\vt^q)=
-\r_3 \ {}^{(1)}\!A_{pq}*(\vt^p\wedge\vt^q).
\end{equation} 
The terms (\ref{term1}) and (\ref{term2}) are algebraic independent, thus 
the 2-form $\S$ vanishes if anf only if $\r_2=\r_3=0$.
\eprf\\
The 2-form $\S$ is a diffeomorphic and a global $SO(1,3)$ invariant. 
Certainly, it is not transforms invariantly under local  $SO(1,3)$ 
transformations of the coframe field. 
However, the Lagrangian itself does not have such invariance for a 
generic choice of the parameters. 
So, for all values of parameters (excepting the case of the teleparallel 
equivalent of GR) the  2-form $\S$ is a well defined object. 
The antisymmetric part of the energy-momentum tensor is known 
by Poincar{\'e} gauge theory of gravity \cite{hehl95} to    
represent the rotational (spin) properties of the field. 
The corresponding 2-form $\S$ of the electromagnetic field vanishes identically. 
Also the 2-form $\S$ being calculated for the Schwarzschild solution 
is zero \cite{itin2}. 
A rotational solution of the general free parametric coframe  field equation  may 
produce an example of a model with a non-zero  2-form $\S$. 
%------------------------
\subsect{Antisymmetric property of the current} 
%------------------------
To complete the consideration let us examine the possibility to have a 
model with a current corresponding to a pure antisymmetric tensor. 
Certainly, such model (if it exists) can not be related to a some 
viable physical situation because it have to describe a 
non-trivial dynamics with zero energy.
Via (\ref{ant}) the current corresponding  to a pure antisymmetric tensor 
have to satisfy the relation 
$\T_a=\frac 12 \ \vt_a\wedge \S.$ 
\pro 
The symmetric part of a tensor corresponding to the coframe current (\ref{cur}) 
in non-zero (in general) for all values of the parameters $\rho_i$,  
i.e., for all coframe models.
\epro
\prf
The coframe current (\ref{cur}) is traceless for arbitrary  values
of the parameters $\r_1,\r_2,\r_3$. 
Thus, by (\ref{pr1-6}), the symmetric part of the tensor satisfy 
${}^{(sym)}\T_a=T_{(ab)}*\vt^b$. 
Consequently, it is zero if and only if 
$${}^{(sym)}\T_a=\T_a-\frac 12 \vt_a\wedge\S=0$$
or using (\ref{cur})
\begin{equation}\label{eqq0}
(e_a\hook\C_m)\wedge *\F^m-e_a\hook L-\frac 12 \vt_a\wedge\S=0.
\end{equation} 
Let us represent the LHS of this equation via the global 
$SO(1,3)$ covariants \cite{i-k}. 
All  terms of (\ref{eqq0}) are linear in the parameters 
$\rho_i$. Thus, the contributions corresponding to different parameters 
may be computed separately.    \\
Calculate the first term of (\ref{eqq0}). 
The $\rho_1$-contribution takes the form 
 (the notation $\vt^{ab\cdots}=\vt^a\wedge\vt^b\wedge\cdots$ is used)
\br\label{eqq1}
&&\r_1(e_a\hook\C^m)\wedge *\C_m=
\frac 12\r_1 \ {C^m}_{aq}C_{mrs}\vt_q\wedge *(\vt^{rs})=-\r_1{{C^m}_a}^qC_{mqs}*\vt^s
\nonumber\\
&&=\r_1\! \  {}^{(6)}\! \! A_{ab}*\vt^b
\er
The $\rho_2$-contribution is 
\br\label{eqq2}
&&\r_2(e_a\hook\C^m)\wedge *\Big(e_m\hook(\vt^n\wedge\C_n)\Big)=
\frac 12 \r_2{C^m}_{aq}C_{mrs}\vt^q\wedge*(\d^n_m\vt^{rs}-2\d^r_m\vt^{ns}) \nonumber\\
&&=-\r_2{{C^m}_a}^q\Big(C_{mqs}+C_{qsm}+C_{smq}\Big)*\vt^s \nonumber\\
&&=
\r_2\Big({}^{(6)}\! \! A_{ab}-{}^{(5)}\! \! A_{ab}-{}^{(4)}\! \! A_{ab}\Big)*\vt^b
\er
The $\rho_3$-contribution is 
\br\label{eqq3}
&&\r_3(e_a\hook\C^m)\wedge *\Big(e_n\hook(\vt^m\wedge\C^n)=
\frac 12\r_3 C_{mpq}{C^n}_{rs}(e_a\hook\vt^{pq})\wedge*(e_n\hook\vt^{mrs})\nonumber\\
&&=-\r_3\Big(C_{maq}{C^{mq}}_s-{C^m}_{am}{C^n}_{ns}+{C_{sa}}^q{C^n}_{nq}\Big)*\vt^s\nonumber\\
&&=\r_3\Big({}^{(6)}\! \! A_{ab}-{}^{(7)}\! \! A_{ab}-{}^{(1)}\! \! A_{ab}\Big)*\vt^b
\er
Summing (\ref{eqq1}), (\ref{eqq2}), (\ref{eqq3}) we obtain the  first term of 
(\ref{eqq0}). 
As for a contribution of the second term of (\ref{eqq0}) it is of the form: 
\begin{equation}\label{eqq4}
e_a\hook L=-(\eta_{ab}*L)\wedge*\vt^b.
\end{equation}
This term is expressed by scalar valued invariants (\ref{2-20})-(\ref{2-22}).\\
The contribution of the third term of (\ref{eqq0}) is represented by (\ref{term1}) and 
(\ref{term2}) as  
\br\label{eqq5}
\vt_a\wedge\S&=&\vt_a\wedge\Big(\r_2 \ {}^{(4)}\!A_{pq}+\r_2 \ {}^{(5)}\!A_{pq}-\r_3 \ {}^{(1)}\!A_{pq}\Big)*(\vt^p\wedge\vt^q)\nonumber\\
&=&\Big(\r_3 \ {}^{(1)}\!A_{[ab]}-\r_2 \ {}^{(4)}\!A_{[ab]}\Big)*\vt^b
\er
Recall, we are looking for which  values of the parameters the LHS of (\ref{eqq0}) 
is vanished identically. 
Using the algebraic independence of the covariants ${}^{(i)}\! A_{ab}$ we obtain that 
$\r_2$ $\r_3$ should vanish in order to illuminate  the contribution of 
${}^{(5)}\! A_{ab}$ and ${}^{(7)}\! A_{ab}$, correspondingly. 
Now, $\r_1$ have also to be zero in order to illuminate  the contribution of 
${}^{(1)}\! A_{ab}$. 
Consequently, the coframe current can not be pure antisymmetric in any 
of coframe models.
\eprf
%------------------------
\sect{Symmetric reduction of the field equation}
%------------------------
The  2-form $\S$ and consequently the antisymmetric part of the current 
are not vanished in all viable models. 
Thus, the field equation (\ref{feq1}) represents a system of 16 independent PDE. 
In the case $2\rho_2+\rho_3\ne 0$ it is a well determined system for 
16 independent components of the coframe. \\
As for the case $2\rho_2+\rho_3=0$ (the teleparallel equivalent of GR) 
the situation is rather different. 
The corresponding Lagrangian accepts an additional invariant transformation - 
local Lorentz transformation  of the coframe field. 
Such invariance of the Lagrangian certainly preserves on the field equation level. 
The coframe variable, however, has only 10 independent components, 
related to the components of the metric tensor. 
Consider what is the situation with the field equation. 
Also here we are dealing with pure coframe field in vacuum, i.e., in (\ref{feq1}) the 
energy-momentum current of the material field ${}^{(mat)}\T_a$ is taken to be zero.  
Rewrite the field equation (\ref{feq1}) in the form 
\begin{equation}\label{eq1}
\E_a:=d*\F_a- \T^a=0.
\end{equation} 
The vector-valued 3-form $\E_a$ accepts the decomposition to the 
symmetric and anti- \\symmetric parts 
\begin{equation}\label{eq2}
\E_a={}^{(sym)}\E_a+{}^{(ant)}\E_a.
\end{equation} 
Similarly to the current  $\T_a$ the antisymmetric part represents as 
\begin{equation}\label{eq3}
{}^{(ant)}\E_a=\vt_a\wedge \E,
\end{equation} 
where $\E:=e_a\hook \E^a$ is a scalar-valued 2-form. 
Using the coframe invariant notations (see Appendix): 
 we express the 2-form $\E$ as \cite{i-k}
\br\label{4-5}
\E&=&\Big(-2(\rho_1-2\rho_2-\rho_3) ({}^{(1)}\!B_{[ab]}+{}^{(1)}\!A_{[ab]})+
\nonumber\\
&&\qquad (2\rho_2+\rho_3) \ ({}^{(3)}\!B_{[ab]}+
{}^{(2)}\!A_{[ab]})\Big)*(\vt^a\wedge\vt^b).
\er
The RHS of this equation is identically zero if and only if
\begin{equation}\label{4-6}
\rho_1=0,  \qquad 2\rho_2+\rho_3=0.
\end{equation}
This is the case of the teleparallel equivalent of the 
Einsteinian gravity. 
Consequently,  the system of 16 field equations is restricted to  the symmetric 
system of 10 independent equations. 
Thus, it represents a well defined system of PDE. 
This result shows also that  the Einstein equation (in the coframe versus) is the 
unique symmetric field equation that can be derived from the 
 quadratic coframe Lagrangian. 
%----------------------------------
\sect{Concluding remarks}
%----------------------------------
We consider a class of coframe models determined by the values of three 
dimensionless free parameters. 
In the case $\rho_1\ne 0$ the field equation  has no a spherical 
symmetric solution with Newtonian behavior at infinity. 
However, in the case $\rho_1=0$ all the models have the same Schwarzschild 
solution for arbitrary values of the remaining parameters. 
Consequently all these models can be considered as viable \cite {Hehl98}, 
\cite{Blag2}. 
The most of viable coframe Lagrangians (with $\r_2+2\r_3\ne 0$) represent 
gravity models alternative to GR. 
A local conservative 3-form of energy-momentum current is well defined 
for such models. 
This object preserves all the invariance transformations of the corresponding 
Lagrangian. 
The tensor-form correspondence produces a diffeomorphic invariant and 
global Lorentz covariant energy-momentum tensor of  coframe field. 
This tensor is traceless for all values of parameters. 
Consequently, the gravitons in quantum extensions of all such models have to be 
massless.\\ 
The antisymmetric part of the tensor corresponds to an invariant 2-form 
$\S$, which is related to the rotation properties of the field. 
The tensor is proved to have 
in general non vanishing symmetric and antisymmetric parts in all viable 
models. \\
The exceptional case $\r_2+2\r_3= 0$ describes a coframe model with additional 
local Lorentz invariance of the Lagrangian. 
It is an alternative coframe (teleparallel) description of GR, not an alternative 
gravity model.  
In this case the coframe is defined only up to local 
pseudo-rotations. 
The corresponding system of field equations is restricted to a system 
of 10 independent PDE for 10 independent components of the coframe. 
Thus, it is a well determined system. \\
The conserved current, however, does not preserves  local 
Lorentz transformations, as in  standard GR. \\
Two possibilities is open in this situation.\\
The first topic is to study the coframe models as an
alternative to GR.  The known obstacle to this is the particle
analysis \cite{S-N}, \cite{kuh}, which shows the existence of non-physical modes
(ghosts, tachyons).  We plan to examine if these modes appear also in
the coframe models in a way which is slightly different from the Einstein gravity.\\ 
The second topic is to consider  standard GR as a limit of the 
free parametric teleparallel model with $\r_2+2\r_3\to 0$. Such a limit produces 
one more symmetry of the Lagrangian.  It is the local
(pointwise) pseudo-rotations of the coframe.  This symmetry is
transported to the field equations.  Unfortunately the metric
construction of GR prevents $\T^a$ to share this property.  Still the
proper defined integrals of the current may be invariant.  Also
meaningful asymptotic invariants may be properly defined by the coframe
current.
\section*{Acknowledgments}
I am deeply grateful to F.W. Hehl and to S. Kaniel  for useful discussion, valuable  suggestions and 
comments. 

\appendix
%----------------------------------
\sect{Coframe invariants}
%----------------------------------
We present a list of $SO(1,3)$ covariants \cite{i-k}. 
All of them are invariant under the 
diffeomorphisms of the manifold. 
The first order derivative covariant is defined as 
\begin{equation}\label{2-3a}
{C^a}_{bc}=e_c\hook(e_b\hook d\vt^a).
\end{equation} 
The contraction of this 3-indexed object with Lorentz metric produces a 1-indexed 
object 
\begin{equation}\label{2-3b}
C_a={C^m}_{ma}=e_a\hook(e_m\hook d\vt^m).
\end{equation} 
This first order derivative $SO(1,3)$ covariant is proportional to the coderivative of the 
coframe $d^\dagger\vt^a=*d*\vt^a$.\\
The second order derivatives of the coframe are expressed in $SO(1,3)$ covariant form 
via a 4-indexed object: 
\begin{equation}\label{2-4b}
{B^a}_{bcd}=e_d\hook{C^a}_{bc} =e_d\hook d \Big(e_c\hook(e_b\hook d\vt^a)\Big).
\end{equation}
This 4-indexed covariant object is a coframe analog of the Riemannian curvature tensor. 
The field equation can involve only 2-indexed object. Three possible 
contractions of ${B^a}_{bcd}$: 
\begin{eqnarray}
\label{2-5}
{}^{(1)}B_{ab}&=&{B_{abm}}^m,\\
\label{2-6}
{}^{(2)}B_{ab}&=&{B^m}_{mab},\\
\label{2-7}
{}^{(3)}B_{ab}&=&{B^m}_{abm}
\er
are coframe analogs to the Ricci curvature tensor.\\
The unique scalar valued second order invariant is
\begin{equation}\label{2-9}
B={{B^a}_{ab}}^b= {B^a}_{abc}\eta^{bc}={}^{(1)}{B^a}_a={}^{(2)}{B^a}_a.
\end{equation}
This is an analog to the Riemannian curvature scalar.\\
The field equation as well as the energy-momentum tensor of the coframe field 
also involve terms quadratic in the first order derivatives.  
The possible two-indexed $SO(1,3)$ covariants quadratic in first order derivatives are:
\br 
\label{2-13} %{A1}
{}^{(1)}\!A_{ab}&:=&C_{abm}C^m, \qquad {}\\
\label{2-14} %{A2}
{}^{(2)}\!A_{ab}&:=&C_{mab}C^m \qquad {\rm\ \ antisymmetric \ object,}\\
\label{2-15} %{A3}
{}^{(3)}\!A_{ab}&:=&C_{amn}{C_b}^{mn}\qquad{\rm symmetric \ object,}\\
\label{2-16} %{A4}
{}^{(4)}\!A_{ab}&:=&C_{amn}{{C^m}_b}^n, \qquad {}\\
\label{2-17} %{A5}
{}^{(5)}\!A_{ab}&:=&C_{man}{{C^n}_b}^m\qquad{\rm symmetric \ object,}\\
\label{2-18} %{A6}
{}^{(6)}\!A_{ab}&:=&C_{man}{{C^m}_b}^n\qquad{\rm symmetric \ object,}\\
\label{2-19} %{A7}
{}^{(7)}\!A_{ab}&:=&C_aC_b\qquad{\rm\ \ \ \  \ \ \ symmetric \ object.}
\er
%}\end{equation}
In addition to the 2-indexed $A$-objects the general field equation may also 
include their traces multiplied by $\eta_{ab}$.
These traces of  2-indexed objects are  scalar $SO(1,3)$ invariants:
\br
\label{2-20} %{A11}
{}^{(1)}\!A&:=& {}^{(1)}{A_a}^a=-{}^{(7)}{A_a}^a,\\
\label{2-21} %{A22}
{}^{(2)}\!A&:=& {}^{(3)}{A_a}^a={}^{(6)}{A_a}^a,\\
\label{2-22} %{A33}
{}^{(3)}\!A&:=& {}^{(4)}{A_a}^a={}^{(5)}{A_a}^a.
\er
Three scalars ${}^{(i)}\!A$ constitute three independent parts of the 
coframe Lagrangian. 

%\begin{references}
%{999}

%\end{eqnarray}
\end{document}